\documentclass[twocolumn,prl,showpacs,preprintnumbers,amsmath,amssymb]{revtex4}

\usepackage{epsfig}
\usepackage{graphicx}
\usepackage{dcolumn}
\usepackage{bm}
\usepackage{wasysym}
\usepackage{marvosym}
\usepackage{amsmath,amssymb}
\usepackage{color}

\makeatletter

\usepackage{ulem}
\newcommand{\simleq}{\mbox{\raisebox{-1.0ex}{$\stackrel{<}{\sim}$}}}

\makeatother
\begin{document}

\title{Biased bilayer graphene: semiconductor with a gap tunable by electric
  field effect}
\author{Eduardo V. Castro$^1$, K. S. Novoselov$^2$, S. V. Morozov$^2$,
N.~M.~R.~Peres$^3$, J.~M.~B.~Lopes~dos~Santos$^1$, Johan Nilsson$^4$, F.
Guinea$^{5}$, A. K. Geim$^2$, and  A. H. Castro Neto$^{4,6}$}

\affiliation{$^1$CFP and Departamento de F\'{\i}sica,
  Faculdade de Ci\^encias Universidade do Porto, P-4169-007 Porto, Portugal}

\affiliation{$^2$Department of Physics and Astronomy, University of Manchester,
  Manchester, M13~9PL, UK}

\affiliation{$^3$Center of Physics and Departamento de
  F\'{\i}sica, Universidade do Minho, P-4710-057 Braga, Portugal}

\affiliation{$^4$Department of Physics, Boston University, 590
  Commonwealth Avenue, Boston, MA 02215,USA}

\affiliation{$^5$Instituto de Ciencia de Materiales de
  Madrid. CSIC. Cantoblanco. E-28049 Madrid, Spain}

\affiliation{$^6$Department of Physics, Harvard University, Cambridge, MA
  02138, USA}


\begin{abstract}
We demonstrate that the electronic gap of a graphene bilayer can be
controlled externally by applying a gate bias. From the
magneto-transport data (Shubnikov-de Haas measurements of the
cyclotron mass), and using a tight binding model, we extract the
value of the  gap as a function of the electronic density. We show
that the gap can be changed from zero to mid-infrared energies
by using fields of $\simleq 1$ V/nm, below the electric breakdown
of SiO$_2$. The opening of a gap is clearly seen in the quantum Hall
regime.
\end{abstract}


\maketitle

The electronic structure of materials is given by their chemical
composition and specific arrangements of atoms in a crystal lattice
and, accordingly, can be changed only slightly by external factors
such as temperature or high pressure. In this Letter we show, both
experimentally and theoretically, that the band structure of bilayer
graphene can be controlled by an applied electric field so that the
electronic gap between the valence and conduction bands can be tuned
between zero and mid-infrared energies. This makes
bilayer graphene the only known semiconductor with a tunable energy
gap and may open the way for developing photodetectors and lasers
tunable by the electric field effect. The development of a
graphene-based tunable semiconductor being reported here, as well as
the discovery of anomalous integer quantum Hall effects (QHE) in
single layer~\cite{NGM+05,ZTS+05} and unbiased bilayer~\cite{NMcCM+06} 
graphene, which are associated with massless~\cite{PGN06} and 
massive~\cite{MF06} Dirac fermions, respectively,
demonstrate the potential of these systems for carbon-based
electronics~\cite{johan}. Furthermore, the deep connection between
the electronic properties of graphene and certain theories in
particle physics makes graphene a test bed for many ideas in basic
science.

Below we report the experimental realization of a tunable-gap
graphene bilayer and provide its theoretical description in terms of
a tight-binding model corrected by charging effects (Hartree
approach)~\cite{edmc}. Our main findings are as follows: ({\it i})
in a magnetic field, a pronounced plateau at zero Hall conductivity
$\sigma_ {xy}=0$ is found for the biased bilayer, which is absent in
the unbiased case and can only be understood as due to the opening
of a sizable gap, $\Delta_g$, between the valence and conductance
bands; ({\it ii}) the cyclotron mass, $m_\textrm{c}$, in the bilayer
biased by chemical doping  is an asymmetric function of carrier
density, $n$, which provides a clear signature of a gap and allows
its estimate; ({\it iii}) by comparing the observed behavior with
our tight-binding results, we show that the gap can be tuned to
values larger than 0.2~eV; ({\it iv}) 
we have crosschecked our theory against  
angle-resolved photoemission spectroscopy (ARPES) data~\cite{arpes}
 and found excellent agreement.

The devices used in our experiments were made from bilayer graphene
prepared by micromechanical cleavage of graphite on top of an
oxidized silicon wafer (300~nm of SiO$_2$)~\cite{Netal04_short}. By
using electron-beam lithography, the graphene samples were then
processed into Hall bar devices similar to those reported in 
Refs.~\cite{NGM+05,ZTS+05,NMcCM+06}. To induce charge carriers, we applied
a gate voltage $V_g$ between the sample and the Si wafer, which
resulted in carrier concentrations $n_1 = \alpha V_g$ due to the
electric field effect. The coefficient $\alpha \cong
7.2\times10^{10}\,\textrm{cm}^{-2}/\textrm{V}$ is determined by the
geometry of the resulting capacitor and is in agreement with the
values of $n_1$ found experimentally~\cite{NGM+05,ZTS+05,NMcCM+06}.
In order to control independently the gap value and the
Fermi level  $E_{\textrm{F}}$, the devices could also be doped
chemically by exposing them to NH$_3$ that adsorbed on graphene and
effectively acted as a top gate providing a fixed electron density
$n_0$~\cite{doping_science}. 
The total bilayer density $n$ is then 
$n = n_1 + n_0$ relatively to half-filling. 
The electrical measurements were carried out
by the standard lock-in technique in magnetic fields up to 12T and
at temperatures between 4 and 300K.

\begin{figure}
\begin{center}\includegraphics*[width=0.98\columnwidth]{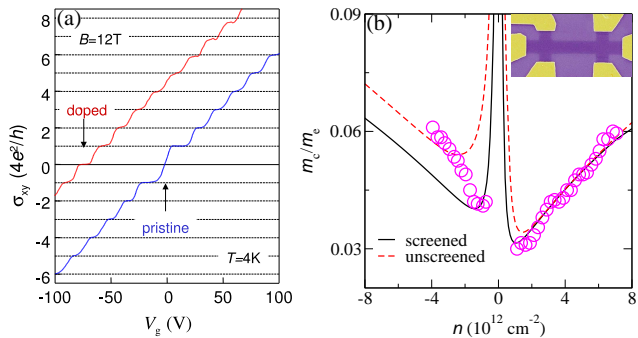}
\end{center}
\caption{\label{cap:expHall}(color online) 
(a)~Measured Hall conductivity of pristine (undoped) and chemically doped
bilayer graphene ($n_0 \approx 5.4\times10^{12}\,\textrm{cm}^{-2}$), 
showing a comparison of the QHE in both systems.
(b)~Cyclotron mass vs $n$,
normalized to the free electron mass, $m_\textrm{e}$.
Experimental data are shown as $\ocircle$. The solid line is the
result of the self-consistent procedure and the dashed
line corresponds to the unscreened case. The inset shows an electron
micrograph (in false color) of one of our Hall bar devices with a
graphene ribbon width of 1~$\mu$m.}
\end{figure}

We start by showing experimental evidence for the gap opening in
bilayer graphene. 
Figure~\ref{cap:expHall}(a) shows the measured Hall conductivity of 
bilayer graphene, which allows a comparison of the QHE behavior in 
the biased and unbiased cases.
Here the curve labeled
``pristine'' shows the anomalous QHE that is characteristic of the
unbiased bilayer~\cite{NMcCM+06}. In this case, the Hall
conductivity exhibits a sequence of plateaus at
$\sigma_{xy}=4Ne^{2}/h$ where $N$ is integer and the factor 4 takes
into account graphene's quadruple degeneracy. The $N=0$ plateau is
strikingly absent, so that a double step of $8e^{2}/h$ in height
occurs at $n=0$, indicating a metallic state at the neutrality 
point~\cite{NMcCM+06}. Note that the back-gate voltage induces asymmetry
between the two layers but QHE measurements can only probe states
close to $E_{\textrm{F}}$ and are not sensitive to
the presence (or absence) of a gap below the Fermi sea. To probe the
gap that is expected to open at finite $V_g$, we first biased the
bilayer devices chemically and then swept 
$V_g$ through the
neutrality point, in which case $E_\textrm{F}$ passes between the
valence and conduction bands at high $V_g$. The energy gap is
revealed by the appearance of the $N=0$ plateau at $\sigma_{xy}=0$
[see the curve labeled ``doped'' in Fig.~\ref{cap:expHall}(a)]. The
emerged plateau was accompanied by a huge peak in longitudinal
resistivity $\rho_{xx}$, indicating an insulating state (in the
biased device, $\rho_{xx}$ at $n=0$ exceeded 150 kOhm at 4~K, as
compared to $\approx 6$ kOhm for the unbiased case under the same
conditions). The recovered sequence of equidistant plateaus
represents the "standard" integer QHE that would be expected for an
ambipolar semiconductor with an energy gap exceeding the cyclotron
energy. The latter is estimated to be $>40$ meV in the case of
Fig.~\ref{cap:expHall}(a).

To gain further information about the observed gap, we measured the
cyclotron mass of charge carriers and its dependence on $n$. To this
end, we followed the same time-consuming procedure as described in
detail in Ref.~\cite{NGM+05} for the case of single-layer graphene.
In brief, for many different gate voltages, we measured the
temperature ($T$) dependence of Shubnikov-de Haas oscillations and
then fitted their amplitude by the standard expression
$T/\sinh(2\pi^2k_\textrm{B}Tm_\textrm{c}/\hbar e B)$. To access
electronic properties of both electrons and holes in the same
chemically biased device, we chose to dope it to $n_0 \approx
1.8\times10^{12}\,\textrm{cm}^{-2}$, i.e. less than in the case of
Fig. 1. The results are shown in Fig.~\ref{cap:expHall}(b). The linear
increase of $m_\textrm{c}$ with $|n|$ and the pronounced asymmetry
between hole- and electron-doping of the biased bilayer are clearly
seen here.

\begin{figure}
\begin{center}\includegraphics*[width=0.98\columnwidth]{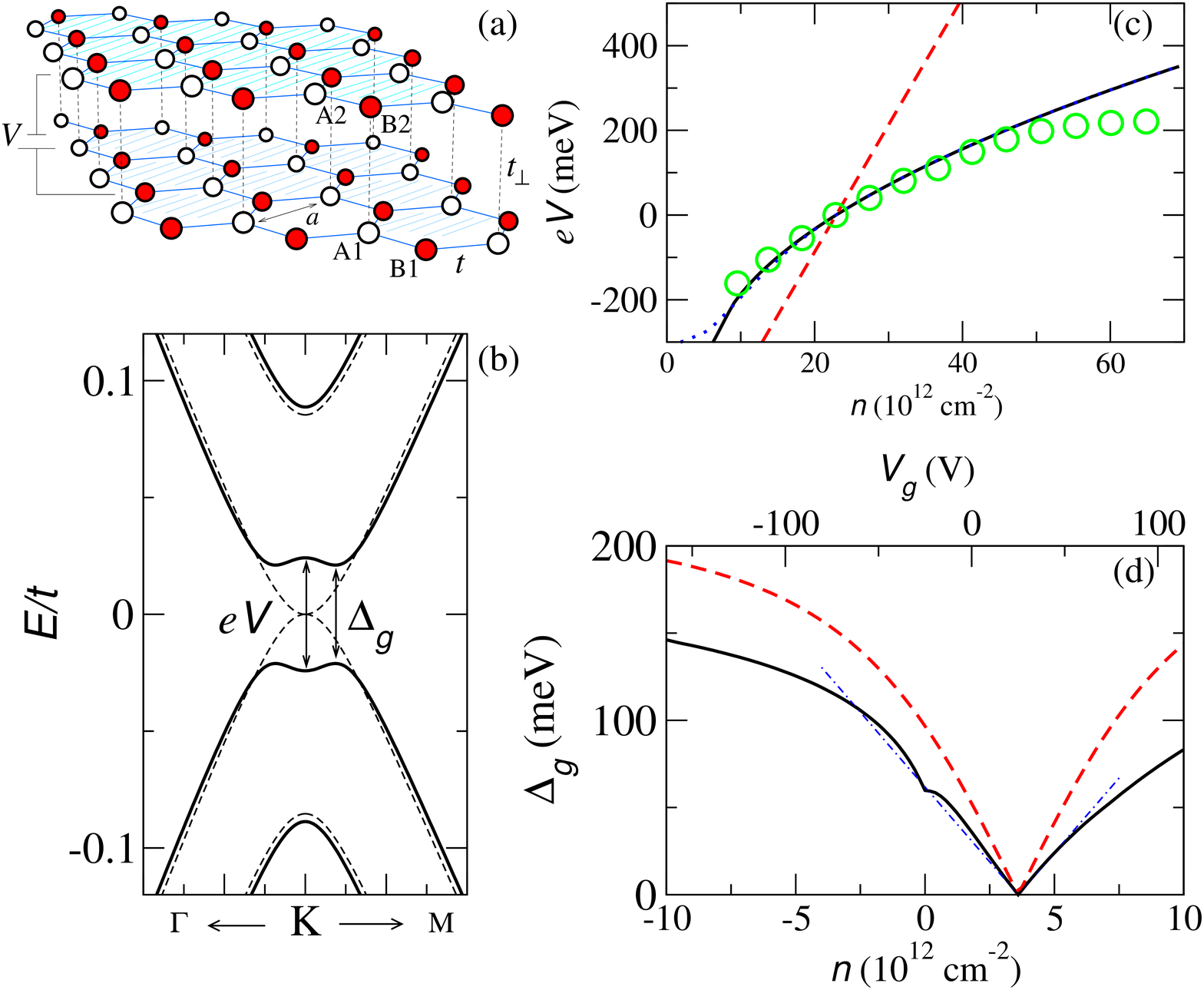}
\end{center}
\caption{\label{cap:gap}(color online) (a)~Lattice structure of
bilayer graphene and parameters of our model (see text). (b)~Band
structure of bilayer graphene near the Dirac points for $eV=150$~meV
(solid line) and $V=0$ (dashed). (c)~$eV$ as a function of
 $n$: solid and dotted lines are the result of the
self-consistent procedure (see text) for $t_\perp = 0.2$~eV and
$t_\perp = 0.4$~eV, respectively; dashed line is the unscreened
result; circles represent $eV$ vs. $n$ measured by ARPES~\cite{arpes}.
(d)~Band gap as a function of $n$ (bottom axis) and $V_g$ (top):
solid and dashed lines are for the screened and unscreened cases,
respectively. The thin dashed-dotted line is a linear fit to the
screened result at small biases.}
\end{figure}

To explain the
observed Hall conductivity and cyclotron mass data for bilayer
graphene, in what follows, we shall use a tight-binding description
of electrons in bilayer graphene.
Its carbon atoms are arranged in two honeycomb lattices labeled~1 and~2
and stacked according to the Bernal order ($A$1-$B$2), where~$A$ and~$B$
refer to each sublattice within each honeycomb layer, as shown in
Fig.~\ref{cap:gap}~(a). The system is parameterized by a
tight-binding model where $\bm\pi$-electrons are allowed to hop
between nearest-neighbor sites, with in-plane hopping $t$ and
inter-plane hopping $t_\perp$. Throughout the Letter we use
$t=3.1$~eV and $t_\perp=0.22$~eV. The value of $t$ is inferred from
the Fermi-Dirac velocity in graphene, $v_\textrm{F} =
(\sqrt{2}/3)at/\hbar \approx 10^6 \textrm{ms$^{-1}$}$, where $a
\approx 2.46\,\textrm{\r{A}}$ is the same-sublattice carbon-carbon
distance, and $t_{\perp}$ is extracted by fitting $m_\textrm{c}$
(see below). For the biased system the two layers gain different
electrostatic potentials, and the corresponding energy difference is
given by $eV$. The presence of a perpendicular magnetic field
$\mathbf{B} = B\,\textrm{\^e}_z$ is accounted for through the
standard Peierls substitution, $t \rightarrow t \exp\{i e
\smallint_{\mathbf{R}}^{\mathbf{R}+\bm\delta}
\mathbf{A}\cdot\textrm{d}\mathbf{r} \}$, where $e$ is the electron
charge, $\bm \delta$ the vector connecting nearest-neighbor sites,
and $\mathbf{A}$ the vector potential (in units such that
$c=1=\hbar$).

Fig.~\ref{cap:gap}~(b) shows the electronic structure of the biased
bilayer near the Dirac points ($K$ or~$K'$).
In agreement with the Hall conductivity results in Fig.~\ref{cap:expHall}(a), 
one can see that the unbiased gapless semiconductor (dashed line) becomes, 
with the application of an electrostatic potential $V$, a small-gap 
semiconductor (solid line) whose gap is given by:
$ \Delta_g= [e^2 V^2 t_{\perp}^{2}/(t_{\perp}^{2} + e^2 V^2)]^{1/2} $. 
As $V$
can be externally 
controlled, 
this model predicts that biased bilayer graphene should be a tunable-gap 
semiconductor, in agreement with results obtained previously using a 
continuum model~\cite{edmc}.
Note that the gap does not reach a minimum at the~$K$ point due to 
the ``mexican-hat'' dispersion at low energies~\cite{stack}.

The electric field induced between the two layers can be considered
as a result of the effect of charged surfaces placed above and below
bilayer graphene . Below is an accumulation or depletion layer inthe Si wafer, which has charge density $n_1e$. Dopants above the
bilayer effectively provide the second charged surface with density
$n_0e$. Assuming equal charge $-ne/2$ in layers~1 and~2 of the
bilayer we find an {\it unscreened} potential difference given by,
\begin{equation}
\label{eq:Vns}
V = (2 - n/n_0) \, n_0 e d/(2\varepsilon_0) \, ,
\end{equation}
where $\varepsilon_0$ is the permittivity of free space, and
$d\cong0.34$~nm is the interlayer distance. A more realistic
description should account for the charge redistribution due to the
presence of the external electric field. For given~$V$ and $n$, we
can estimate the induced charge imbalance between layers $\Delta
n(n,V)$ through the weight of the wave functions in each layer
(Hartree approach; also, see~\cite{edmc}). This charge imbalance is
responsible for an internal electric field that screens the external
one, and a self-consistent procedure to determine the {\it screened}
electrostatic difference requires 
\begin{equation}
\label{eq:Vs}
V = (2 - n/n_0 + \Delta n(n,V)/n_0) \,
n_0 e d/(2\varepsilon_0) \, .
\end{equation}
Zero potential difference and zero gap  are expected at $n=2n_0$ in both unscreened and
screened cases, as seen from Eqs.~(\ref{eq:Vns}) and~(\ref{eq:Vs})
and the fact that $\Delta n(n,0)=0$.

In Fig.~\ref{cap:gap}~(c) our calculations using Eqs.~(\ref{eq:Vns})
and~(\ref{eq:Vs}) are compared with  
ARPES measurements of the $V$ dependence on $n$ in bilayer graphene
by~Ohta {\it et al}.~\cite{arpes}. In their experiment, $n$-type doping with
$n_1^\textrm{ex} \approx 10\times 10^{12}\,\textrm{cm}^{-2}$ was due
to the SiC substrate and therefore fixed. The electronic density
$n_0$ induced by the deposition of K atoms onto the vacuum side was
then used to vary the total density. A zero gap was found around
$n\approx 23\times 10^{12}\,\textrm{cm}^{-2}$ from which value we
expect $n_1^\textrm{th} \approx 11\times 10^{12}\,\textrm{cm}^{-2}$,
in agreement with the experiment. In order to compare the behavior
of $V$  with varying $n$ we replace $n_0$ in
Eqs.~(\ref{eq:Vns}) and~(\ref{eq:Vs}) with $n_0 = n -
n_1^\textrm{th}$. The result for the unscreened case
[Eq.~(\ref{eq:Vns})]~--~shown in Fig.~\ref{cap:gap}~(c) as a dashed
line~--~cannot describe the experimental data. The solid and dotted
lines are the screened results obtained with the self-consistent procedure
[Eq.~(\ref{eq:Vs})] for $t_\perp = 0.2$~eV and $t_\perp = 0.4$~eV,
respectively; both are in good agreement with the experiment, except
in the gap saturation
regime 
 at $n\gtrsim 50 \times 10^{12}\,\textrm{cm}^{-2}$.

For the experiments described in the present work, the expected
behavior of the gap with varying $n$ or, equivalently, $V_g$ is
shown in Fig.~\ref{cap:gap}~(d). The dashed line is the unscreened
result [$V$ given by Eq.~(\ref{eq:Vns})] and the solid line is the
screened one [Eq.~(\ref{eq:Vs})]. In both cases, the chemical doping
was set to $n_0 = 1.8\times 10^{12}\,\textrm{cm}^{-2}$ at which
$m_\textrm{c}$ was measured in our experiment (equivalent of
$V_g \approx 25$~V). The dashed-dotted (blue) line is a linear fit to
the screened result for small gap, yielding
$\Delta_g(\textrm{meV}) = \beta \vert V_g(\textrm{V}) - 25\vert$
with a coefficient $\beta \approx 1.2~\textrm{meV}/\textrm{V}$. The
linear fit is valid in the small-gap regime ($\Delta_g \ll t_\perp$)
only, and the theory predicts a gap saturation to $\Delta_g \sim
t_\perp$ at large biases. Note that the breakdown field for SiO$_2$
is $1$ V/nm (i.e. 300~V for the used oxide thickness) and,
therefore, practically the whole range of allowed gaps (up to
$t_\perp$) should be achievable for the demonstrated devices.

To explain the observed behavior of 
the cyclotron mass, $m_\textrm{c}$,
shown in Fig.~\ref{cap:expHall}(b),
 we used the
semi-classical expression $m_\textrm{c}(n) = (\hbar^2/2\pi)\partial
A(E)/\partial E|_{E=E_\textrm{F}(n)}$, where $A(E)$ is the $k$-space
area enclosed by the orbit of energy $E$ and $n$ the carrier density
at $E_{\textrm{F}}$. In Fig.~\ref{cap:expHall}(b) our theory results are
shown as dashed and solid lines for the unscreened and screened
description of the gap, respectively (analytical expressions for
$m_\textrm{c}$ in the biased bilayer will be given 
elsewhere~\cite{next}). The inter-layer coupling $t_\perp$ is the only
adjustable parameter, as $t$ is fixed and $V$ is given by
Eq.~(\ref{eq:Vns}) or Eq.~(\ref{eq:Vs}). The value of $t_\perp$
could then be chosen so that theory and experiment gave the same
$m_\textrm{c}$ for $n\approx 3.6\times 10^{12} \textrm{cm}^{-2}$. At
 this particular density the gap closes  and the theoretical value becomes independent of the
screening assumptions. We found $t_\perp \approx 0.22$~eV, in good
agreement with values found in the literature. The theoretical
dependence $m_\textrm{c}(n)$ agrees well with the experimental data
for the case of electron doping. Also, as seen in Fig.~\ref{cap:expHall}(b),
the screened result provides a somewhat better fit than the
unscreened model, especially at low electron densities. This fact,
along with the good agreement found for the potential difference data of
Ref.~\cite{arpes} [see Fig.~\ref{cap:gap}~(c)], allows us to conclude that for doping of the same
sign from both sides of bilayer graphene, the gap is well described
by the screened approach. In the hole doping region in Fig.~\ref{cap:expHall}(b), the Hartree approach underestimates the value of
$m_\textrm{c}$ whereas the simple unscreened result overestimates
it. This can be attributed to the fact that the Hartree theory used
here is reliable only if the gap is small compared to $t_{\perp}$.
In our experimental case, $n_0>0$ and, therefore, the theory works
well for a wide range of electron doping $n>0$, whereas even a
modest overall hole doping $n <0$ corresponds to a significant
electrostatic difference between the two graphene layers. In this
case, the unscreened theory overestimates the gap whereas the
Hartree calculation underestimates it. However, it is clear that the
experimental data
in Fig.~\ref{cap:expHall}(b)
 interpolate between the screened result at low
hole doping and the unscreened one for high hole densities. This
indicates that the true gap actually lies between the unscreened and
screened limits [see Fig.~\ref{cap:gap}~(d)], and that a more
accurate treatment of screening is needed when $eV$ becomes of the
order of $t_{\perp}$.

\begin{figure}
\begin{center}\includegraphics*[%
  scale=0.3]{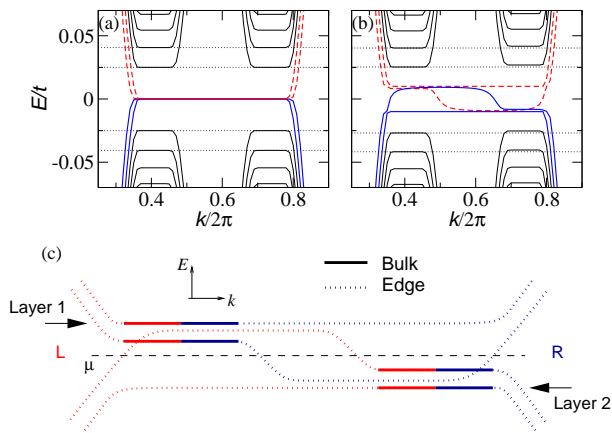}\end{center}
\caption{\label{cap:LLV}(color online)
 Energy spectrum for a ribbon of bilayer graphene with zigzag edges, 
$t_{\perp}/t=0.2$, \textbf{$B=30\,$}T, and width $N=400$
unit cells: (a)~$eV=0$;  (b)~$eV=t_{\perp}/10$. (c)~Sketch of the bands
close to zero energy (for the biased bilayer) with indication of
bulk (solid lines) or edge (dotted lines) states and their left~($L$) or right~($R$)
positions along the ribbon.
Quasidegeneracies have been removed for clarity.}
\end{figure}

In what follows, 
we model and discuss the QHE data presented in Fig.~\ref{cap:expHall}(a). 
We consider a ribbon of bilayer graphene~\cite{note}
with zigzag edges (armchair edges give similar results).
Fig.~\ref{cap:LLV} shows the energy spectrum in the presence of a
strong magnetic field. Panel~(a) corresponds to the unbiased case 
[see the curve labeled ``pristine'', Fig.~\ref{cap:expHall}(a)],
where the four degenerate bands at zero energy contain four
degenerate bulk Landau levels~\cite{MF06} and four surface states
characteristic of the bilayer with zigzag edges~\cite{next}. The
spectrum for a biased device is shown in panel~(b). In this case two
flat bands with energies $-eV/2$ and $eV/2$ appear, similar to the
case of zero magnetic field. The  other two zero energy bands become
dispersive inside the gap, showing the band-crossing phenomenon. The
Landau level spacing is set by $\gamma \equiv \sqrt{3/2}ta/l_B$
($l_B$ is the magnetic length), and as long as $eV\ll t_{\perp}$, the
bias is much smaller than the Landau level spacing at low fields.
Then non-zero Landau levels in the bulk are almost insensitive to
$V$, as seen in Fig.~\ref{cap:LLV}~(b), except for a small asymmetry
between Dirac points. A close inspection of Fig.~\ref{cap:LLV} shows
that the valley degeneracy is lifted due to the different nature of
the Landau states at  $K$ and $K'$ valleys with respect to their
projection in each layer. The valley asymmetry has a stronger effect
in the zero energy Landau levels, where the charge imbalance is
saturated. This opens a gap of $eV$ in size. Also, there is an
intra-valley degeneracy lifting [Fig.~\ref{cap:LLV}~(c)], because
only one of the two Landau states of the unbiased system remains an eigenstate
when a bias is applied. For $eV\gtrsim t_{\perp}$ (not shown in
Fig.~\ref{cap:LLV}) the dispersive modes start crossing with
non-zero bulk Landau levels.

Let us now 
model the measured
Hall conductivity for the biased bilayer graphene,
which is shown in the (red) curve labeled ``doped'' in Fig.~\ref{cap:expHall}(a).
We consider the case of the chemical potential lying
inside the gap, between the last hole- and the first electron- like
bulk Landau levels, and crossing the dispersive bands as shown in
Fig.~\ref{cap:LLV}~(c). As pointed out by Laughlin~\cite{Laugh81},
changing the magnetic flux through the ribbon loop by a flux quantum
makes the states to move rigidly towards one edge. In the usual
integer QHE, the energy increase due to this adiabatic flux
variation results in the net transfer of $n\times g$ electrons (spin
and valley degeneracy $g$) from one edge to the other, and the
quantization of the Hall conductivity follows the expression~\cite{Halp82}:
 ${\cal I}/{\cal V}=gne^{2}/h$, where ${\cal I}$ is
the current carried around the loop and ${\cal V}$ the potential
drop between the two edges. However, in the present case there is no
net charge transfer across the ribbon. As seen in Fig.~\ref{cap:LLV}
(c), the band states at the chemical potential belonging to the same
band are surface states localized at the same edge (see the figure
caption for details). The rigid movement of the states towards one
edge makes an electron-hole pair to appear at both edges, resulting
in zero net charge transfer. Therefore, we expect a Hall plateau
$\sigma_{xy}=0$ showing up when the carriers change sign, i.e. at
the neutrality point. Accordingly, the Hall conductivity of the
biased bilayer is given by $\sigma_{xy}=4 N e^{2}/h$ for all integer
$N$, including zero. Note that at the zero Hall plateau the current
carried around the ribbon loop is zero, ${\cal I}=0$, which implies,
from the theory view point, a diverging longitudinal resistivity at
low $T$, in stark contrast to all the other Hall plateaus that
exhibit zero $\rho_{xx}$, as in the standard QHE. This behavior has
been observed experimentally, as discussed above with reference to
Fig.~\ref{cap:expHall}(a). This concludes our interpretation of the
experimental data.

E.V.C., N.M.R.P., and J.M.B.L.S. were supported by
POCI 2010 via project PTDC/FIS/64404/2006 and FCT through 
Grant No.~SFRH/BD/13182/2003.
F.G. was supported by MEC (Spain) grant No.~FIS2005-05478-C02-01 and EU
contract 12881 (NEST).
A.H.C.N was supported through NSF grant DMR-0343790.
The experimental work was supported by EPSRC (UK).

\bibliographystyle{apsrev}

\begin{thebibliography}{43}

\expandafter\ifx\csname natexlab\endcsname\relax\def\natexlab#1{#1}\fi
\expandafter\ifx\csname bibnamefont\endcsname\relax
  \def\bibnamefont#1{#1}\fi
\expandafter\ifx\csname bibfnamefont\endcsname\relax
  \def\bibfnamefont#1{#1}\fi
\expandafter\ifx\csname citenamefont\endcsname\relax
  \def\citenamefont#1{#1}\fi
\expandafter\ifx\csname url\endcsname\relax
  \def\url#1{\texttt{#1}}\fi
\expandafter\ifx\csname urlprefix\endcsname\relax\def\urlprefix{URL }\fi
\providecommand{\bibinfo}[2]{#2}
\providecommand{\eprint}[2][]{\url{#2}}

\bibitem{NGM+05}
K. S. Novoselov {\it et al.}, Nature {\bf 438}, 197 (2005).

\bibitem{ZTS+05}
Y. Zhang {\it et al.}, Nature {\bf 438}, 201 (2005).

\bibitem{NMcCM+06}
K. S. Novoselov {\it et al.}, Nature Phys. {\bf 2}, 177 (2006).

\bibitem{PGN06}
\bibinfo{author}{\bibfnamefont{N.~M.~R.} \bibnamefont{Peres}},
  \bibinfo{author}{\bibfnamefont{F.}~\bibnamefont{{G}uinea}}, \bibnamefont{and}
  \bibinfo{author}{\bibfnamefont{A.~H.} \bibnamefont{{C}astro {N}eto}},
  \bibinfo{journal}{Phys. {R}ev. {B}} \textbf{\bibinfo{volume}{73}},
  \bibinfo{pages}{125411} (\bibinfo{year}{2006}).

\bibitem{MF06}
\bibinfo{author}{\bibfnamefont{E.}~\bibnamefont{Mc{C}ann}} \bibnamefont{and}
  \bibinfo{author}{\bibfnamefont{V.~I.} \bibnamefont{{F}al'ko}},
  \bibinfo{journal}{Phys. {R}ev. {L}ett.} \textbf{\bibinfo{volume}{96}},
  \bibinfo{pages}{086805} (\bibinfo{year}{2006}).

\bibitem{johan}
  A. K. Geim, K. S. Novoselov, Nature Mat. {\bf 6}, 183 (2007); J. Nilsson {\it et al.}, cond-mat/0607343.

\bibitem{edmc}
Theory of the gap opening within the continuum model was previously
given by E. McCann, Phys. Rev. B {\bf 74}, 161403 (2006).

\bibitem{arpes}
  T.~Ohta {\it et al.},
  Science {\bf 313}, 951 (2006).



\bibitem{Netal04_short} K. S. Novoselov {\it et al.}, Science {\bf 306},
666 (2004).

\bibitem{doping_science}F. Schedin {\it et al.}, Nature Mat. {\bf 6}, 652 (2007).


\bibitem{stack}
F. Guinea, A. H. Castro Neto, and N. M. R. Peres, Phys. Rev. B {\bf 73},
245426 (2006).


\bibitem{next}
E. V. Castro {\it et al.}, unpublished.

\bibitem{note}The unbiased case, in the continuum approximation,
was studied in Ref.~\cite{MF06}.


\bibitem{Laugh81}
\bibinfo{author}{\bibfnamefont{R.~B.} \bibnamefont{Laughlin}},
  \bibinfo{journal}{Phys. {R}ev. {B}} \textbf{\bibinfo{volume}{23}},
  \bibinfo{pages}{5632} (\bibinfo{year}{1981}).

\bibitem{Halp82}
\bibinfo{author}{\bibfnamefont{B.~I.} \bibnamefont{Halperin}},
  \bibinfo{journal}{Phys. {R}ev. {B}} \textbf{\bibinfo{volume}{25}},
  \bibinfo{pages}{2185} (\bibinfo{year}{1982}).



\end{thebibliography}

\end{document}